# Dispersion relation, propagation length and mode conversion of surface plasmon polaritons in silver double-nanowire systems

**Shulin Sun,[1,2] Hung-Ting Chen,[2] Wei-Jin Zheng,[2] and Guang-Yu Guo[2,3,*]**

[1]*National Center for Theoretical Sciences, Physics Division, National Taiwan University, Taipei 10617, Taiwan*
[2]*Department of Physics, National Taiwan University, Taipei 10617, Taiwan*
[3]*Graduate Institute of Applied Physics, National Chengchi University, Taipei 11605, Taiwan*

[*]*gyguo@phys.ntu.edu.tw*

**Abstract:** We study the surface plasmon modes in a silver double-nanowire system by employing the eigenmode analysis approach based on the finite element method. Calculated dispersion relations, surface charge distributions, field patterns and propagation lengths of ten lowest energy plasmon modes in the system are presented. These ten modes are categorized into three groups because they are found to originate from the monopole-monopole, dipole-dipole and quadrupole-quadrupole hybridizations between the two wires, respectively. Interestingly, in addition to the well studied gap mode (mode 1), the other mode from group 1 which is a symmetrically coupled charge mode (mode 2) is found to have a larger group velocity and a longer propagation length than mode 1, suggesting mode 2 to be another potential signal transporter for plasmonic circuits. Scenarios to efficiently excite (inject) group 1 modes in the two-wire system and also to convert mode 2 (mode 1) to mode 1 (mode 2) are demonstrated by numerical simulations.

## 1. Introduction

Surface plasmon polaritons (SPPs), a kind of electromagnetic excitations coupled with the collective oscillation of conduction electrons at the metal/dielectric interface [1], receive much attention recently in the physics, chemistry and biology communities. Thanks to their sub-wavelength and local field enhancement effects, SPPs can be applied in diverse areas [2, 3], such as the near-field microscopy [4], bio- and chemical- sensors [5], enhanced Raman [6, 7] and nonlinear optics [8-10]. Depending on the dimension of the metallic systems, the supported SPP modes can be either the localized or propagating ones. The localized surface plasmon resonances are very sensitive to the geometry and material type of the nanostructures. The localized plasmon excitations in various metallic nanoparticles such as spheres, ellipsoids, shells and cylinders, have been extensively studied [11-12]. Except for the single particle, the coupling effects in multi-particle systems are hot topics since they can give rise to fascinating physical properties [13]. Furthermore, the propagating SPP modes can serve as the information carriers in such systems as particle chains [14-16], metallic grove structures [17] and coupling nanowire arrays [18-22], which are good candidates to meet the needs of the fast optical nanodevices and nanocircuits.

In this work, we focus on the metallic nanowire systems which have already attained keen interests in the SPP waveguides [18-22], resonators [23-25], and lasers [26]. It is well known that an intrinsic problem to restrict the applications of surface plasmons is the energy loss. To overcome this obstacle, hybrid surface plasmon modes supported by a composite waveguide of metal, spacer and dielectric, have been introduced [26-28]. A large amount of energy is driven away from the metal leading to less energy loss and longer propagation length for the hybrid plasmon modes. Here the thickness of the spacer is an important factor [26-28]. Indeed, much attention has been paid to the coupling between the pure SPP modes in the nanostructures with same/different geometries and materials types [14-22]. For example, a strongly interacting nanowire array could serve as a robust plasmon waveguide [18] in which the gap SPP modes were introduced as an excellent candidate to transport signals over tens of micrometers in the near-infrared region. Furthermore, they are robust against the variation of the wire cross section and the curvature which is advantageous for the technological applications [18]. To further enhance the signal transfer rate, the coupling effects of the gap plasmons in the multi-wire waveguides with the different configurations and intermediate wires were also studied [19]. To selectively excite certain SPP modes in metallic nanowire pair systems, the trajectory of an electron beam was used [20]. In [18-21], the boundary element method [29] was employed to calculate the electron energy loss probability by these systems. The energy loss spectra may reveal the dispersion relations of the excited surface plasmon modes. In more complex coupled systems, the bands may overlap and thus are hardly distinguished from each other [18-21]. In the non-retarded limit, the dispersion relations in single and coupled cylindrical pore systems have been studied. The one-plasmon energy loss probabilities for electron trajectories parallel to the cylindrical axis were also calculated using a Hamitonian formulism [30]. Furthermore, the plasmonic resonant coupling between metallic wires for both non-touching and intersecting configurations was also investigated by the plane wave excitation, and a dramatic field enhancement between the wires in a small separation was observed [31]. Nevertheless, despite of the intensive investigations mentioned above, the detailed relations between the plasmon modes in the single and double-wire systems have not been systematically examined. Experimentally, limited by the excitation method or mode strength, some SPP modes may be either too difficult to excite or hardly visible. In such situations, one may have to turn to other numerical methods to overcome these problems.

Here we systematically investigate dispersion relations, surface charge distributions, field patterns and propagation lengths of the ten lowest energy plasmon modes supported by a metallic double-nanowire system by using the eigenmode analysis based on the finite element method (FEM). We find these ten modes to originate from the monopole-monopole, dipole-dipole and quadrupole-quadrupole hybridizations between the two wires, and thus categorize them into three groups, respectively. We show that, in addition to the well known gap mode

(mode 1), the other mode from group 1 has a larger group velocity and a longer propagation length than mode 1, indicating mode 2 to be another potential signal transporter for plasmonic circuits. Furthermore, based on our numerical simulations, we propose methods to efficiently inject group 1 modes in the two-wire system and also to convert mode 2 (mode 1) to mode 1 (mode 2).

## 2. Computational method

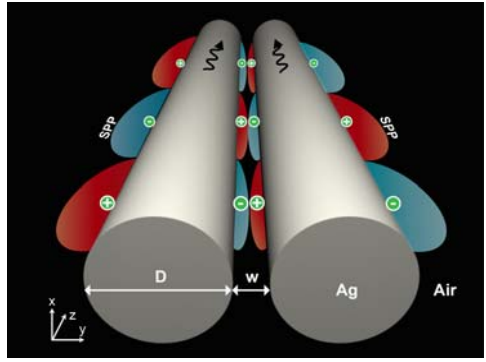

Fig. 1. Illustration of the infinite long silver double-nanowire system studied here and also the supported hybridized SPP modes. Here $D$ and $w$ denote the wire diameter and the separation of the two nanowires, respectively.

The FEM-based eigenmode analysis is a powerful approach to study photonic crystals and metamaterials [32, 33] where the propagating wave modes are the main interests. Here we apply this method to study the SPP modes in a silver double-nanowire system. The FEM method implemented in the COMSOL Multiphysics® software is used. The energy bands, surface charge distributions, field patterns are efficiently calculated using this approach. Based on this information, ten lowest energy SPP modes are investigated and their hybridizations are analyzed. Furthermore, the propagation lengths of these ten SPP modes are determined to evaluate their propagation properties as possible signal transporters.

The infinite long silver double-nanowire system considered here is illustrated in Fig.1. The wire diameter $D$ is 200 nm and the wire-wire distance $w$ will be varied to study the gap size effects. For simplicity, the surrounding medium is chosen to be air. The permittivity of silver is described by $\varepsilon = 1 - \omega_p^2 / [\omega(\omega + i\Gamma)]$ with $\omega_p = 13.7 \times 10^{16}$ Hz and $\Gamma = 2.73 \times 10^{13}$ Hz determined through fitting the experimental data [34]. In our simulations, the nanowires are put in a large simulation box of more than ten wavelength size, surrounded by the absorbing boundaries. The box size must be large enough to prevent the absorbing boundaries from influencing the localized SPP modes. The eigenvalue solver is used to calculate the effective index $n_{eff} = n' + in''$ of the SPP modes. Here the effective index $n_{eff}$ at a given frequency $\omega$ is determined by an iterative calculation, and the obtained $n_{eff}$ is used as the guess value for the next frequency. A high mesh density and a large box size are used to ensure the calculated $n_{eff}(\omega)$ converged with respect to both the mesh density and the box size. The real and imaginary parts of the eigen-moments of the SPP modes are then derived from $k_z = n'k_0$ and $k_z^i = n''k_0$. And the propagation length is given by $L = 1/(2k_z^i)$ [18, 25].

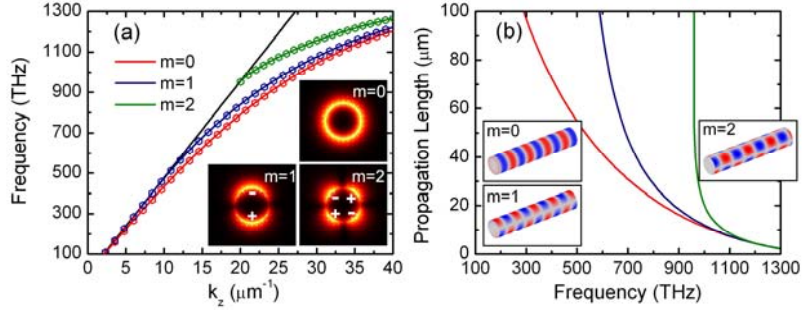

Fig. 2. (a) Dispersion relations and (b) propagation lengths of the SPP modes in the single silver nanowire, obtained via the eigenmode analysis based the FEM simulations. Here $k_z$ is the real part of the momentum of the SPP modes along the nanowire. Modes $m = 0$, 1, and 2 are monopole, dipole and quadrupole SPP modes, respectively. In (a), the dispersion relations from the Mie theory [Eq. (4)] are also displayed as hexagon dots. In the insets in (a) and (b), the electric field patterns and surface charge distributions are shown, respectively. In the insets in (b), the red and blue colors denote the positive and negative surface charges, respectively.

As a benchmark, an infinite single silver nanowire is studied first and the results will be also used to understand the double-nanowire system. The dispersion relations and the propagation lengths of the $m = 0$, 1, and 2 SPP modes in the single silver nanowire are shown as the solid lines in Figs. 2(a) and 2(b), respectively. Here $m$ denotes the angular quantum number and the mode fields in a cylindrical system have the form of $\sim e^{im\phi}$ where $\phi$ is the azimuthal angle. The electric field patterns and the surface charge distributions calculated based on the Gauss law, are also displayed in the insets in Fig. 2. Clearly, the $m = 0$, 1, 2 modes are, respectively, the monopole, dipole and quadrupole modes of the single nanowire. At a fixed frequency, the propagation length increases with the $m$ value (Fig. 2). Furthermore, the real part $k_z$ of the parallel momentum decreases with the increasing $m$. This indicates that the SPP mode with a larger $m$ is less localized in the vicinity of the nanowire and thus has a smaller Ohmic loss, giving rise to an increased propagation length. Interestingly, the propagation length of the $m = 2$ mode becomes extremely large at $\sim 960$ THz. Note that the dispersion relation in this region comes very close to the light line and the SPP modes becomes very delocalized. Therefore, the mode can travel a very long distance, almost like a real propagating wave. Similar behaviors can be found for the other modes.

The SPP modes in the single nanowire can also be obtained analytically from the Mie theory [35]. Here we need to solve the Helmholtz equation $\nabla^2\psi + k^2\psi = 0$ to determine the dispersion relations of the nanowire. The generating function can be expanded by the Bessel and Hankel functions. To avoid a divergent $\psi$ at the wire center $r = 0$, it can only be expressed by the Bessel function of the first kind inside the cylinder. As for the outer space, $\psi$ must be the Hankel function of the first kind since it approaches the propagating wave form of $\sim e^{i\vec{k}_r \cdot \vec{r}}$ when the distance from the center $r$ tends to infinite. The Helmholtz equation can then be written as

$$\begin{cases} \nabla^2\psi^i + k_i^2\psi^i = 0 \\ \nabla^2\psi^o + k_o^2\psi^o = 0 \end{cases} \quad (1)$$

where the wave functions inside and outside the cylinder being, respectively,

$$\begin{cases} \psi^i = \sum_{m=0}^{\infty} C_m J_m(h_i r) e^{im\theta} e^{ik_z z} \\ \psi^o = \sum_{m=0}^{\infty} D_m H_m(h_o r) e^{im\theta} e^{ik_z z} \end{cases}. \quad (2)$$

Here $h_i = \sqrt{k_i^2 - k_z^2}$ and $h_o = \sqrt{k_o^2 - k_z^2}$ are, respectively, the transverse momenta inside and outside the metallic cylinder, and $k_i = \sqrt{\varepsilon^i}\,\frac{\omega}{c}$ and $k_o = \sqrt{\varepsilon^o}\,\frac{\omega}{c}$ are the corresponding total momenta. Continuity condition of the axial momentum and also the angular quantum number on the surface of the cylinder leads to the same $k_z$ and $m$ in the two regions. The electromagnetic fields of the system can be expressed in terms of function $\psi$. By matching the electric and magnetic fields in the two regions on the cylinder surface, we finally get the dispersion relations of the SPP modes in the single metallic nanowire:

$$\left[\frac{1}{h_o}\frac{H'_m(h_o R)}{H_m(h_o R)} - \frac{1}{h_i}\frac{J'_m(h_i R)}{J_m(h_i R)}\right]\left[\frac{k_o^2}{h_o}\frac{H'_m(h_o R)}{H_m(h_o R)} - \frac{k_i^2}{h_i}\frac{J'_m(h_i R)}{J_m(h_i R)}\right] = \frac{m^2 k_z^2}{R^2}\left(\frac{1}{h_i^2} - \frac{1}{h_o^2}\right)^2. \quad (3)$$

In the case of non-radiative surface plasmons, the transverse momenta $h_i$ and $h_o$ should be purely imaginary. Given $h_i \equiv i\kappa_i$ and $h_o \equiv i\kappa_o$, the dispersion relations can be rewritten as

$$\kappa_i^2 \kappa_o^2 \left[\kappa_o \varepsilon^i \frac{I'_m(\kappa_i R)}{I_m(\kappa_i R)} - \kappa_i \varepsilon^o \frac{K'_m(\kappa_o R)}{K_m(\kappa_o R)}\right]\left[\kappa_o \frac{I'_m(\kappa_i R)}{I_m(\kappa_i R)} - \kappa_i \frac{K'_m(\kappa_o R)}{K_m(\kappa_o R)}\right] - \frac{m^2 k_z^2}{R^2}\left(\varepsilon^o - \varepsilon^i\right)^2 \frac{\omega^2}{c^2} = 0.$$
(4)

The dispersion relations for the $m$ =0, 1, 2 modes from Eq. (4) [36] are shown as hexagon dots in Fig. 2. Clearly, these results of the Mie theory agree very well with the FEM simulations, thus serving as a good benchmark of our simulation method.

**3. Results and discussion**

Now let us move on to the silver double-nanowire system first with $w$ = 50 nm. The calculated dispersion relations of the ten lowest energy SPP modes are shown in Fig. 3(a). The propagation lengths of these SPP modes are also shown in Fig. 3(b). The electric field $|\vec{E}|$ patterns for modes 1-10 are displayed in Fig. 4. These modes are named as mode 1 to mode 10 based on their energy levels. The signs of the surface charges determined based on the electric field directions (the white arrows) are also indicated in Fig. 4, and they will be useful for understanding the SPP modes in the double-nanowire system. The dispersion relations of the SPP modes in the metallic double-wire system have been studied before [18-21]. However, the dispersion relations were only implicitly displayed as the peaks in either the electron energy loss probability or photonic density of states spectra in [18-21]. In particular, only the SPP modes excitable by the electron beams would show up in the electron energy loss spectra. Here all the low energy SPP bands are plotted and thus can be clearly identified, although they may be close or even overlap in some frequency regions.

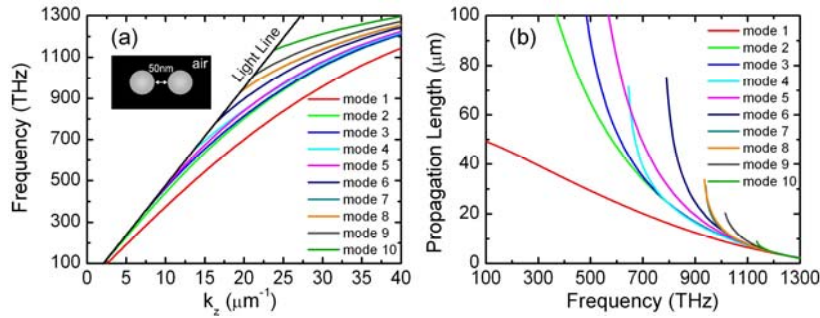

Fig. 3. (a) Dispersion relations and (b) propagation lengths of the ten lowest energy SPP modes on the silver double-nanowire system with $w$ = 50 nm.

In Fig. 2(a), three SPP bands of the $m = 0, 1, 2$ modes in the single nanowire are displayed. Figure 3(a) shows that in the double-wire system, these three SPP bands now split into ten distinguished ones, indicating rather strong hybridizations between the $m = 0, 1, 2$ modes residing on the two different nanowires. Previously, the attention was mainly focused on the gap SPP mode (mode 1) [18-21] because it could transport the signals over tens of micrometers and thus have potential applications in plasmonic nanodevices. Interestingly, Fig. 3(b) shows that the higher order modes could have a much longer propagation length, suggesting that they too could be used as signal transporters. Furthermore, unlike mode 1, the propagation lengths of these higher order modes are rather robust against the wire gap variation, as will be demonstrated below.

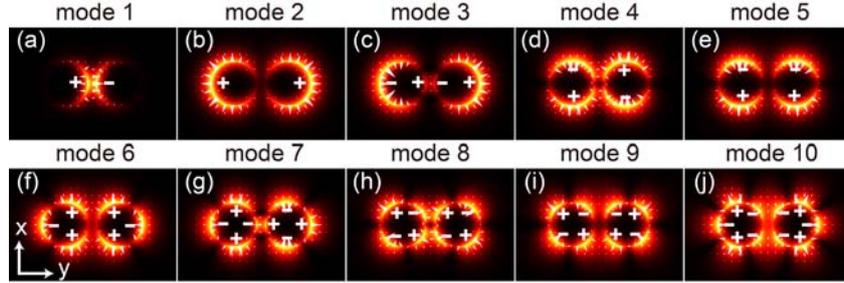

Fig. 4. Electric field patterns of the ten lowest energy SPP modes in the silver double-nanowire system. White arrows denote the directions of electric field. Symbols "+" and "-" indicate the positive and negative surface charges, respectively. The chosen frequency is 1200 THz.

*3.1 Origin and symmetry of the SPP modes*

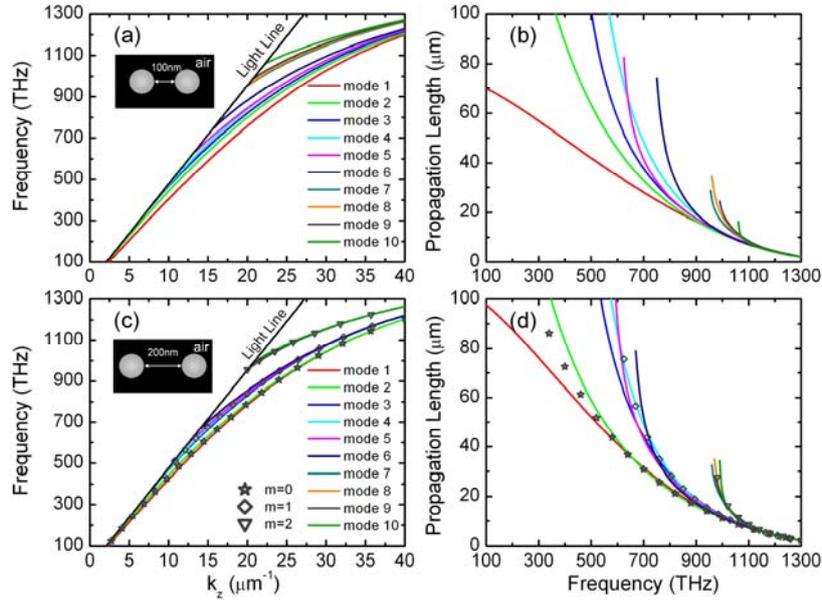

Fig. 5. (a, c) Dispersion relations and (b, d) propagation lengths of the ten SPP modes in the silver double-nanowire system. In (a, b) and (c, d), $w = 100$ and 200 nm, respectively. For comparison, the results of the $m = 0, 1, 2$ SPP modes in the single nanowire are also shown as dots in (c) and (d).

To understand the origins of the ten SPP modes, we increase the distance $w$ of the two nanowires first to 100 nm and then to 200 nm. The results of these calculations are shown in Fig. 5. When the $w$ is increased, the energy bands of the SPP modes get closer and eventually converge to the SPP modes of the single nanowire, because the coupling between the two nanowires becomes weaker. In the $w$ = 200 nm case [Fig. 5(c)], the ten SPP modes can be categorized into three groups: (1) modes 1 and 2; (2) modes 3, 4, 5 and 6; (3) modes 7, 8, 9 and 10. The modes in group 1 and 2 are still distinguishable while the modes in group 3 almost overlap with each other. Since the energy levels of group 3 are high and their relatively short wavelengths thus become comparable with the gap width ($w$ = 200 nm), the coupling between the two nanowires becomes weak. Adding the dispersion relations of the SPP modes ($m$ = 0, 1, 2) in the single nanowire to Fig. 5(c), clearly reveals the correlation of the SPP modes between the single and double nanowire systems. The three groups of the SPP modes in the double-nanowire system clearly originate from the $m$ = 0, 1, 2 modes in the single nanowire, respectively.

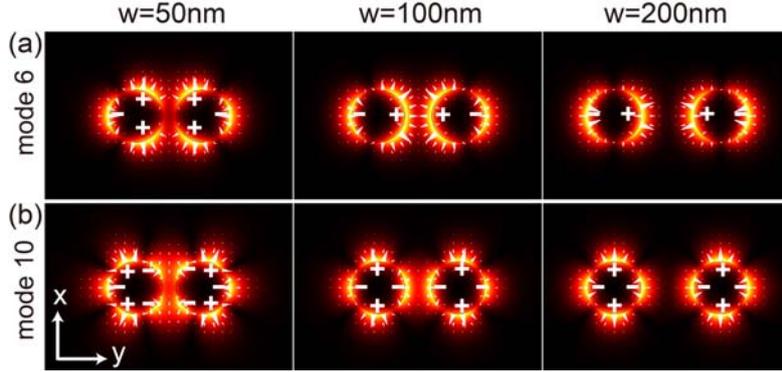

Fig. 6. Evolutions of (a) mode 6 and (b) mode 10 with the increasing gap width $w$. The chosen frequency is 1200 THz.

Surprisingly, Fig. 4 shows that the electric field patterns of modes 6 and 10 have odd numbered petals. That the SPP modes must satisfy the standing wave conditions around the nanowire, would lead to even numbered petals for the field patterns. For a small gap width [Figs. 6(a-b)], however, the like charges on the inner boundaries of the two nanowires would strongly repel each other, and consequently, the charges would be split into two parts in order to reduce the Coulomb repulsion energy. As the gap width is increased, this anomalous behavior disappears because the coupling becomes weaker. All these confirm that modes 6 and 10 are, respectively, the hybridized modes of two dipole and two quadrupole modes on the two individual nanowires in the system.

Based the above analysis, we illustrate the formation of the ten SPP modes due to the hybridizations between the SPP modes on the two nanowires in Fig. 7(a), as in [13]. For group 1 (mode 1 and mode 2), the situation is quite similar to the mode splitting in a metal/insulator/metal structure [37]. The energy level of the symmetric mode (mode 1) is lower than that of the anti-symmetric mode (mode 2) [Fig. 7(a)]. Here the definition of symmetric and anti-symmetric modes is based on their $E_y$ field distributions [37]. The symmetry of mode 1 and mode 2 can be easily seen from the calculated $E_y$ field patterns shown in Fig. 7(d). Clearly, modes 1 and 2 come from the monopole-monopole hybridization. Group 2 (modes 3, 4, 5 and 6) can be further categorized into the collinear and parallel subgroups, depending on the directions of the dipoles on the two nanowires. Obviously the coupling strength of the collinear modes would be stronger because the charges on the two nanowires are closer. Therefore, the splitting of the two collinear modes is larger than the parallel modes, as shown in Fig. 7(b). Since the energy level of the symmetric mode is lower than that of the anti-symmetric one [see Fig. 7 (d)], it can be deduced that the energy level would monotonically increase from mode 3 to mode 6. Clearly, all the group 2 SPP modes originate from the dipole-dipole interactions. A similar analysis would lead to the energy level sequence for the group 3 modes (modes 7-10) shown in Fig. 7(c). They are the products of the quadrupole-quadrupole interactions. Finally, since the symmetry of the infinite double-nanowire system belongs to the point group $D_{2h}$ whose irreducible representations are all one-dimensional, there is no degeneracy for any eigenmodes in our system [38].

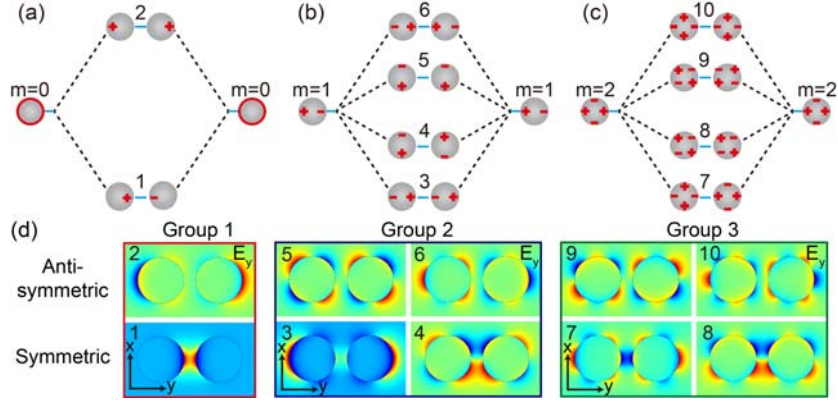

Fig. 7. Classification of the ten hybridized SPP modes in the double-nanowire system into three groups (a-c) based on their origins, namely, (a) group 1 from the monopole-monopole, (b) group 2 from the dipole-dipole, and (c) group 3 from the quadrupole-quadrupole interactions between the two nanowires. The $E_y$ field patterns of the ten modes are shown in (d). Modes 1-10 are denoted by 1-10, respectively. The chosen frequency is 1200 THz

*3.2 Propagation properties*

Now let us focus on the propagation properties of the two lowest SPP modes in the double-nanowire system. Firstly, the propagation length of the gap SPP mode (mode 1) becomes shorter when the gap width is decreased from 200 nm to 100 nm and then 50 nm (Fig. 3 and Fig. 5). This suggests that, to design a gap SPP-based waveguide with a long range propagation, the two nanowires cannot be very close. Furthermore, its propagation length would never be larger than that of the $m = 0$ mode in the single nanowire [Fig. 5(d)]. Interestingly, mode 2 has a larger propagation length than mode 1, as can be seen in Fig. 3 and Fig. 5, indicating that a mode 2-based waveguide may have advantages over the gap SPP-based one.

To further elucidate the important effect of the gap width $w$, we perform simulations for many $w$ values with the wavelength fixed at the telecommunication wavelength $\lambda = 1550$ nm. The calculated propagation lengths of modes 1 and 2 versus the gap width are plotted in Fig. 8(a). Figure 8(a) indicates that a larger gap width would lead to a longer propagation length for mode 1. Nevertheless, the influence of the gap width on mode 2 is much less significant. We also calculate the normalized energy density along the y-direction (the $x = 0$ line) [Figs. 8(c-e)] by adopting the analysis method used in [27]. Here the energy density $w(x, y)$ is defined as [27]:

$$w(x, y) = \frac{1}{2}\left(\frac{d(\varepsilon(x,y)\omega)}{d\omega}|\vec{E}(x,y)|^2 + \mu_0|\vec{H}(x,y)|^2\right), \quad (5)$$

where the diffraction limited area $A_0 = \lambda^2 / 4$ and the total mode energy $W_m$ can be calculated by integrating $w(x, y)$ over the whole simulated region. Interestingly, Figs. 8(c-e) suggest that the normalized energy density $w(x = 0, y)A_0 / W_m$ for mode 1 decreases dramatically with the increasing gap width $w$. This is because for mode 1, most of the electromagnetic energy is stored in the gap region and hence the energy density $w(x, y)$ depends strongly on the gap width $w$. In contrast, the normalized energy density of mode 2 is almost immune to the gap width variation. This is consistent with the nearly constant propagation length of mode 2, as shown in Fig. 8(a). For mode 2, the surface charges have the same sign and are thus mutually repulsive. Consequently, the surface charges are mainly located on the outer boundaries of the two nanowires. Therefore, the coupling of the two SPP modes is weaker since they are more than 400 nm apart. Consequently, mode 2 is less sensitive to the change of the gap width. This explains why the two modes show contrasting behaviors in the propagation length with respect to the gap width variation.

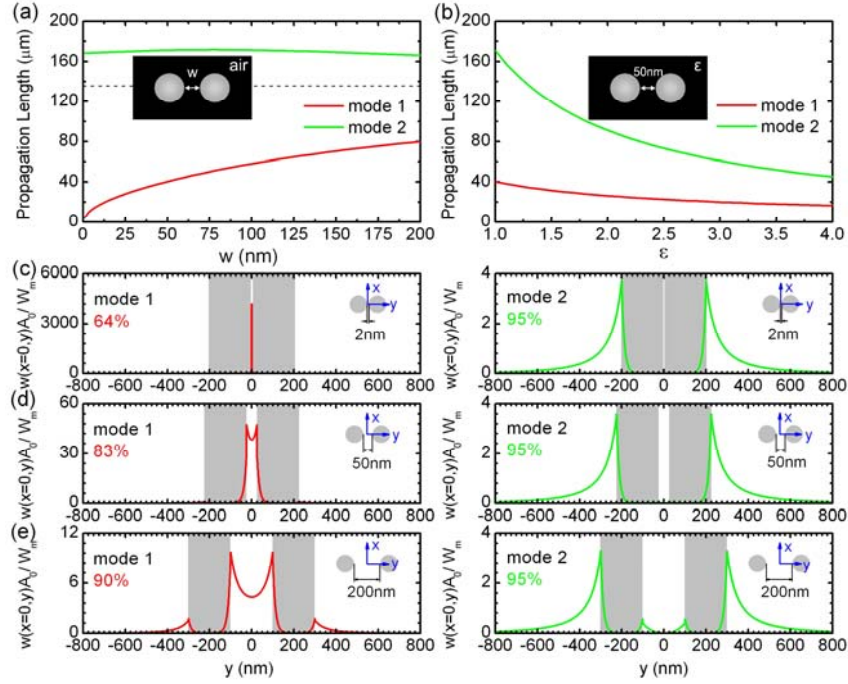

Fig. 8. Propagation lengths of modes 1 and 2 versus (a) the wire gap width $w$ and (b) the permittivity $\varepsilon$ of the surrounding material at the telecommunication wavelength $\lambda = 1550$ nm. In (a), the dashed line denotes the propagation length of the $m = 0$ SPP mode in the single nanowire. The normalized energy densities of modes 1 and 2 along the $x = 0$ line for $w = 2$, 50 and 200 nm are shown in (c-e), respectively. In (c-e), the percentage of the mode energy in the air region is also printed.

The percentage of the mode energy stored in the air region for various gap widths is displayed in Figs. 8(c-e). We can see that for $w = 2$ nm, only about 64 % of the energy in mode 1 is stored and transported in the air, implying a significant energy loss inside the metal. This is why the propagation length of mode 1 is only tens of micrometers [Fig. 8(a)]. As the gap width increases, the percentage in the air region of mode 1 goes up from 64 % to 90 %, resulting in the increased propagation length. For mode 2, in contrast, 95 % of the energy is stored and transported in the outer air region of the two nanowires, and is also independent of the gap width $w$. Therefore, mode 2 would lose much less energy and could travel a much longer distance [Fig. 8(a)]. In fact, the propagation length of mode 2 is as large as ~170 μm, being the same order of magnitude with that reported in [27] (40-150 μm), which, however, originates from a different mechanism.

We also calculate the propagation properties of the double-wire system as a function of the permittivity $\varepsilon$ of the surrounding material to study the effect of the surrounding material. Figure 8(b) shows that, if the permittivity $\varepsilon$ of the surrounding material is increased while the gap width $w$ is fixed at 50 nm, the propagation lengths would decrease. Furthermore, the propagation length of mode 2 would decrease more dramatically than that of mode 1 [Fig. 8(b)]. This can be attributed to the fact that the electromagnetic fields of mode 2 are more delocalized. Therefore, if we want to design a SPP waveguide with a longer propagation range based on the present system, lower refraction index materials are a better choice.

To gain a fuller understanding of the propagation properties of modes 1 and 2, we also study the degree of confinement of these modes. Specifically, we calculate the normalized mode area $A_m/A_0$ for all the ten SPP modes where the mode area $A_m$ is defined as the ratio of the total mode energy to the peak energy density [27]. The inverse of the $A_m/A_0$ of an SPP mode is a good measure of its degree of confinement. Figure 9(a) shows that for all the gap widths considered, mode 1 has a smaller mode area (i.e., a stronger confinement) than the $m = 0$ mode of the single nanowire which, however, has a smaller mode area than mode 2. Figure 9(a) also suggests that the confinement ($A_m/A_0$) of mode 1 becomes weaker (larger) as the gap width $w$ increases, while the degree of confinement of mode 2 is quite insensitive to the gap width variation. Interestingly, the $A_m/A_0$ versus $w$ curves in Fig. 9(a) is very similar to the corresponding propagation length versus $w$ curves in Fig. 8(a), thus indicating a strong correlation between the confinement (mode area) and propagation length. Figure 9(b) shows the degree of confinement of all the ten SPP modes at the frequency of 1200 THz (or $\lambda = 250$ nm) and the gap width $w = 50$ nm. Clearly, the mode area (i.e., mode delocalization) increases nearly monotonically from mode 1 to mode 10. Since the dispersion relations of the ten SPP modes are all far away from the light line and close to each other at 1200 THz [Fig. 3(a)], all the mode areas are small and similar in size [see Fig. 9(b)].

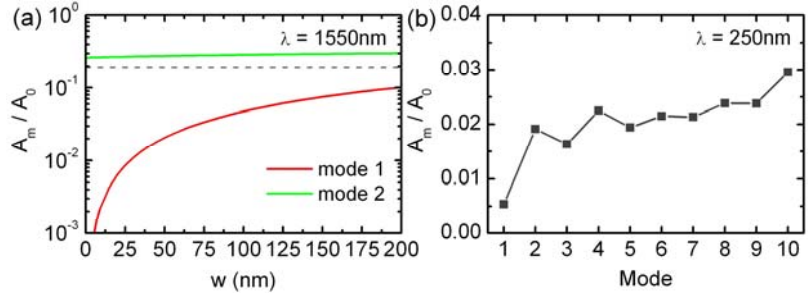

Fig. 9. Normalized mode area $A_m/A_0$ versus (a) gap width $w$ at the telecommunication wavelength $\lambda = 1550$ nm and (b) mode with a fixed gap width $w = 50$ nm at $\lambda = 250$ nm. In (a), the dashed line denotes the $A_m/A_0$ of the $m = 0$ SPP mode in the single silver nanowire.

*3.3 Coupling length of the two neighboring double-nanowire systems*

When the plasmonic waveguides are integrated in a high density plasmonic circuit, the coupling effect between the neighboring units needs to be considered. The coupling length ($L_c$) is defined as $L_c = \pi / |\beta_s - \beta_a|$, where $\beta_s$ and $\beta_a$ are, respectively, the propagation constants of the symmetric and anti-symmetric modes of the two adjacent waveguides. The mode energy in one waveguide would be transferred to another one within a propagation length of $L_c$ [39, 40]. Therefore, the coupling length is an important parameter that determines the integration density of the plasmonic devices. The calculated coupling lengths of modes 1 and 2 for two neighboring double-nanowire systems as a function of their separation $g$ are plotted in Fig. 10. As shown in Fig. 10, the coupling lengths of both mode 1 and 2 increase with the unit separation $g$, because the coupling between the two waveguides becomes weaker. Since the mode energy of mode 1 is mainly stored in the gap region, the coupling effect is relatively weak, thus leading to a large coupling length (Fig. 10).

In addition to the unit separation $g$, the coupling length $L_c$ also depends on how the two waveguides are placed with respect to each other. For example, the field distributions of mode 1 (the gap mode) in the two double-nanowire systems in the on-top geometry [see the inset in Fig. 10(b)] could overlap more strongly than in the side-by-side geometry [see the inset in Fig. 10(a)]. This explains why the coupling length of mode 1 in the former case [Fig. 10(b)] is orders f magnitude smaller in the latter case [Fig. 10(a)] in all the considered $g$ values except at $g \approx 95$ nm where an anomalously large $L_c$ appears. For example, $L_c$ at $g = 200$ nm is ~4100 μm for the side-by-side case and ~80 μm for the on-top case. In the on-top geometry with $g \approx 95$ nm, the propagation constants of the symmetric and anti-symmetric modes would cross each other (i.e., $\beta_s \approx \beta_a$) [41], thus resulting in an extremely large $L_c$ [Fig. 10(b)]. The energy exchange between the two double-nanowire systems is still strong, and thus the mechanism for the large $L_c$ is very different from the non-coupling limit case.

Figure 10 shows that the coupling length for mode 2 is much smaller than that of mode 1. This is because the fields in mode 2 are mainly located on the outer surfaces of the two nanowires, which would make the coupling effect much stronger. At $g = 1000$ nm, the coupling length of mode 2 is ~60 μm in both cases, being smaller than its propagation length. Nevertheless, this coupling length is comparable with that of the slot waveguide pair [42] and the dielectric-loaded plasmonic waveguide [43]. Therefore, from the application view point, mode 2 may offer a good tradeoff between the propagation length and the coupling length [42].

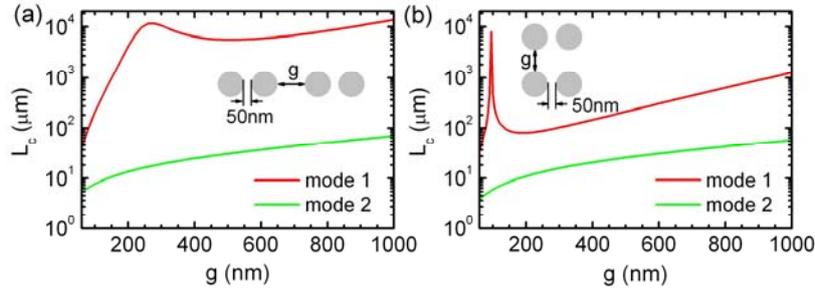

Fig. 10. Coupling length $L_c$ of modes 1 and 2 on two parallel silver double-nanowire systems in (a) a side-by-side (see the inset) geometry and (b) an on-top (see the inset) geometry. Here the gap width $w = 50$ nm and the distance of the two double-nanowire systems is $g$. In (b), $L_c$ becomes singularly large at $g \approx 95$ nm (see the text for explanation).

*3.4 Selective injection of the SPP modes*

Now an interesting question is how to excite the SPP modes in the double-nanowire system to serve as signal transporters. Here we propose three mechanisms to excite mode 2, mode 1 and also their mixed state, respectively. These three mechanisms are demonstrated by the 3D FEM simulations, as illustrated in Fig. 11. For mode 2, the two monopole charges on the two nanowires are coupled in phase. Therefore, mode 2 could be efficiently excited by a *z*-oriented electric dipole source placed at the left end of the two silver nanowires. The field pattern of the excited mode 2 SPP wave is displayed in Fig. 11(b). The SPP wavelength obtained by measuring the periodicity of the simulated $E_y$ field is 1481 nm, being in perfect agreement with the wavelength derived from the calculated dispersion relations shown in Fig. 3. The eigen-momentum $k$ of mode 2 at $\lambda = 1550$ nm is ~4.2 μm$^{-1}$ (Fig. 3), and hence, the SPP wavelength is 1490 nm. If we rotate the dipole away from the *z*-axis in the *x-z* plane, mode 2 could still be excited because the symmetries of the charge pattern of mode 2 and the exciting dipole remains partially matched, although with a much reduced injection efficiency. The relative injection efficiency here is defined as the ratio ($I_{SPP}/I_0$) of the total energy of the SPP wave flowing towards the right side of the absorbing boundary ($I_{SPP}$) for an angle $\theta$ to that for $\theta = 0°$ ($I_0$). Figure 11(a) shows clearly that ratio $I_{SPP}/I_0$ decreases monotonically as angle $\theta$ increases. For example, the electric field $|\vec{E}|$ pattern shown in the inset of Fig. 11(a) indicates that mode 2 could still be excited even at an angle of $\theta = 80°$ away from the *z*-axis, though $I_{SPP}/I_0$ is now reduced by a factor of five. This is because the electric field of the dipole becomes increasingly transverse as angle $\theta$ increases and is completely transverse at $\theta = 90°$. Note that mode 2 is a longitudinal SPP wave.

In contrast, the two monopole charges on the two nanowires in mode 1 are out of phase. Therefore, mode 1 could be efficiently excited only by an electric dipole oriented perpendicular to the nanowire axis (i.e., along the $y$-direction) [see Figs. 11(c) and 11(d)]. The measured SPP wavelength is about 1242 nm based on the simulated $E_y$ field patterns, being again consistent with the value of ~1243 nm derived from the calculated dispersion relation of mode 1 shown in Fig. 3. Figure 11(c) shows that when the dipole is rotated away from the $y$-axis in the $x$-$y$ plane, mode 2 could still be excited but with a decreased injection efficiency. The injection efficiency at $\phi = 80°$ is three times smaller than $\phi = 0°$ (the $y$-axis).

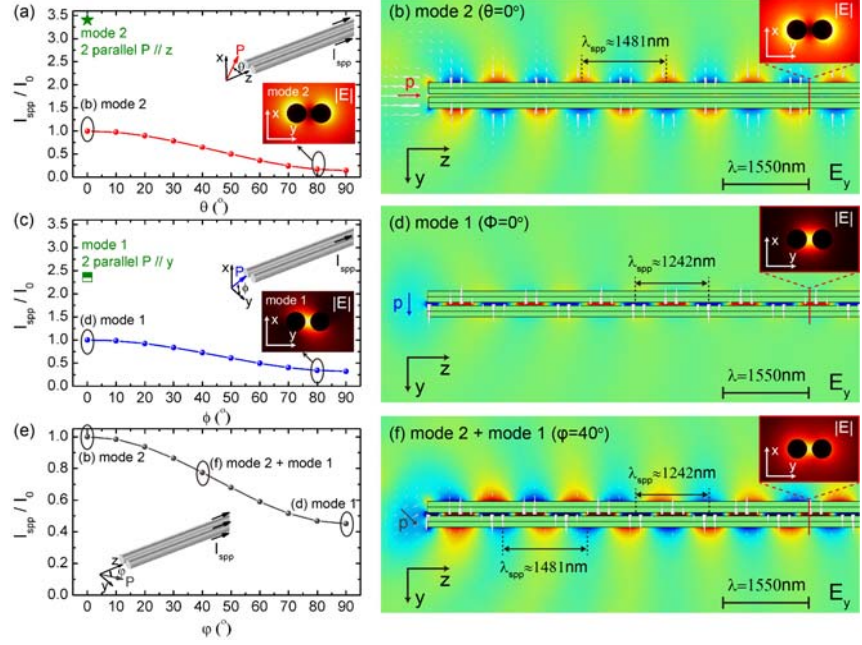

Fig. 11. Excitation of (a, b) mode 2, (c, d) mode 1 and (e, f) their mixed state by a dipole source (denoted by "P") placed at the symmetric point 250 nm away from the left end of the two nanowires. The excited SPP waves are absorbed by the perfect match layer at the right boundary. The total energy flow of the excited SPP modes $I_{SPP}$ is given by an integral over the absorbing boundary. The relative injection efficiency $I_{SPP}/I_0$ as the dipole is rotated in the $x$-$z$ plane (angle $\theta$), the $x$-$y$ plane (angle $\phi$) and the $y$-$z$ plane (angle $\varphi$), is displayed in (a), (c) and (e), respectively. For comparison, the excitation efficiencies $I_{SPP}/I_0$ for injection using two dipoles are also plotted as green symbols in (a, c, e). Two wavelengths $\lambda_{SPP}$ of 1242 and 1481 nm for mode 1 and 2 are measured based on the field patterns in the $y$-$z$ plane displayed in (b, d, f). The white arrows represent the electric fields and the wavelength $\lambda$ = 1550 nm. The electric field $|\vec{E}|$ patterns in the $x$-$y$ plane of the excited SPP modes are shown in the insets in (a-f).

Interestingly, a mixed state of mode 2 and mode 1 could also be excited if the exciting dipole is placed in the $y$-$z$ plane [see Fig. 11(e)]. Furthermore, the percentages of mode 1 and mode 2 may be controlled by varying angle $\varphi$ between the dipole and the $y$-axis. For example, the $E_y$ field pattern at $\varphi = 40°$ [Fig. 11(f)] illustrates that mode 2 and mode 1 are simultaneously excited in the double-nanowire system. The two measured SPP wavelengths are 1242 and 1481 nm, respectively, being consistent with that of mode 1 and mode 2. The electric field $|\vec{E}|$ pattern shown in the inset of Fig. 11(f) appears to be a superposition of that of Figs. 11(b) and 11(d). The energy is stored in both the gap and outer regions of the double-nanowire system. In such an injection process, the two lowest energy SPP modes are simultaneously transported but in the separate spatial regions of the double-nanowire system.

To inject the SPP modes, we could also place a pair of electric dipoles at the end of the two nanowires. In this case, the SPP modes excited in each nanowire could be separately controlled. Furthermore, a specific SPP mode could be injected by using a certain configuration of the dipoles. For example, Figs. 11(a) and 11(c) indicate that a pair of parallel dipoles oriented along the $z$ ($y$) direction could excite a mode 2 (mode 1) SPP wave that is about 3.5 (2.4) times stronger than the corresponding SPP wave excited by one dipole source. Finally, we note that modes 3-6 may also be excited by a pair of dipole sources with appropriate configurations of the dipoles. Nonetheless, a detailed description of this is beyond the scope of this paper.

*3.5 Mode conversion*

The propagation length and field confinement are two important features of a SPP mode. To be a good information carrier, a SPP mode with low field confinement is preferred to reduce the Ohmic loss and to enhance the signal propagation length. On the other hand, if a plasmonic circuit would be placed in a varying dielectric environment, a strongly confined SPP mode may be a better choice because it would be less sensitive to the change of the surrounding mediums. Thus a mode converter for the two fundamental modes in the double-nanowire system will be useful for its diverse applications in plasmonic circuits and light-matter interactions [44, 45].

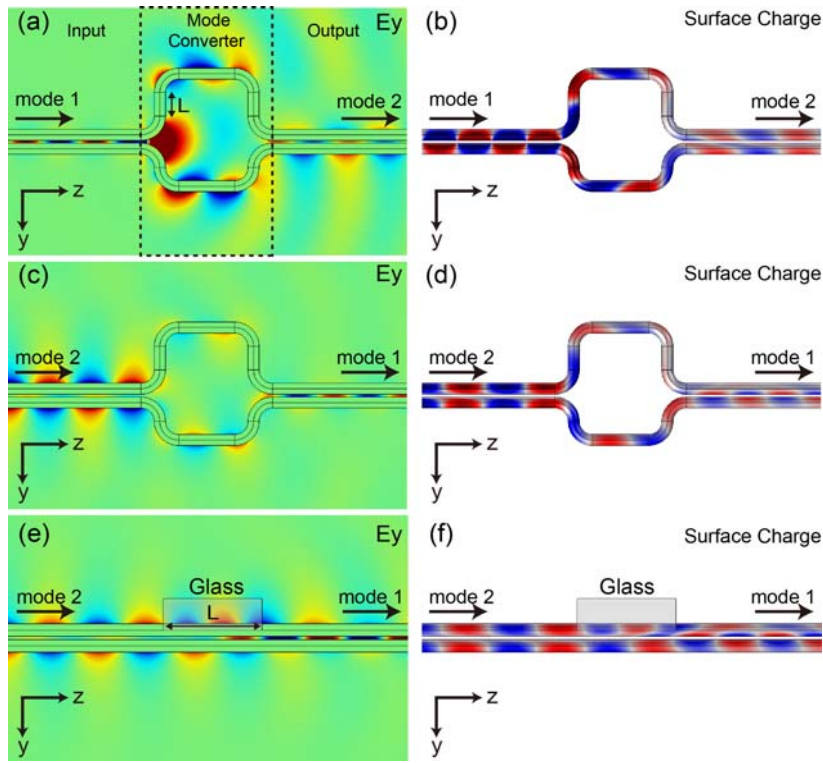

Fig. 12. Proposed mode converters from mode 1 to mode 2 (a-b), and from mode 2 to mode 1 (c-f). The $E_y$ field patterns and surface charge distributions are, respectively, displayed in (a, c, e) and (b, d, f). Here the length $L$ is 430 nm in (a-d) and 1550 nm in (e-f). The refractive index of the glass is 1.5. The wavelength $\lambda$ = 1550 nm. In (a), the color bar range of the $E_y$ field on the left side of the box indicated by the dashed lines is set to be 10 times larger than that on the right side of the box in order to make clear the field patterns of both mode 1 and mode 2.

For example, if a mode 1 SPP wave is to propagate to a detector, the mode 1 SPP wave could be first injected into the double-nanowire system, converted to mode 2 and then transported to a longer distance, and finally converted back to mode 1 as we wish. Such a scenario is illustrated here by a proof-of-principle simulation. As shown in Fig. 12(a), mode 1 is excited at the left side (input) of the plasmonic circuit. The central mode converter then tunes the relative phases of the surface plasmon waves in the two nanowires via tuning the length $L$ of the upper bending nanowire. At the right side (output), mode 1 is converted to mode 2. Unfortunately, as shown in Fig. 12(a), there is some energy loss at the bending area of the circuit and also some small portion of mode 1 survives in the right side region (output) due to robustness of mode 1. Nevertheless, mode 1 is more or less successfully converted to mode 2 by the converter, as demonstrated by the calculated surface charge distribution shown in Fig. 12(b). Clearly, the unlike surface charges at the input region of the circuit become the like ones at the output region due to the transformation by the mode converter. In Figs. 12(c) and 11(d), an inverse conversion from mode 2 to mode 1 by the same system is illustrated. Our simulations prove the scenario in principle, although it could be further optimized by tuning the geometry or employing some other designs for the mode converter.

For mode 1, the coupling between the two nanowires is so strong that we cannot freely manipulate the phase of the SPP wave in one nanowire without affecting the other one. This is why we design a split mode conversion system shown in Figs. 12 (a-d) to convert mode 1 to mode 2. For mode 2, the coupling is much weaker because the field is mainly confined to the outer regions of the two nanowires. Therefore, we could use a simpler system shown in Figs. 12(e-f) to accomplish the mode conversion. Here a block of glass is used to cover the outer surface of one segment of one nanowire and hence the phase of the SPP wave travelling on the nanowire could be controlled by varying the length of the block. Clearly, the energy loss in such a system would be much less than the cases shown in Figs. 12(a-d). However, the inverse conversion from mode 1 to mode 2 is not possible in this system.

## 4. Conclusions

We have carried out a thorough investigation on the SPP modes in the silver double-nanowire system by employing the eigenmode analysis approach based on the FEM simulations. The calculated dispersion relations of the $m = 0, 1, 2$ SPP modes in the single nanowire agree very well with that given by the Mie theory, thereby verifying the reliability of the present approach. Dispersion relations, surface charge distributions, field patterns and propagation lengths of ten lowest energy SPP modes in the silver double-nanowire system are presented. These ten SPP modes are naturally categorized into three groups since they are found to result from the monopole-monopole (2 modes), dipole-dipole (4 modes) and quadrupole-quadrupole (4 modes) interactions, respectively. Group 1 which consists of an antisymmetrically coupled charge mode (i.e., the well studied gap mode) (mode 1) and a symmetrically coupled charge mode (mode 2), are studied in detail. In particular, mode 2 is found to have a larger group velocity and a longer propagation length than the gap mode, and this suggests that mode 2 may be employed as a signal transporter, in addition to mode 1. Scenarios to efficiently excite (inject) these two modes in the two-wire system and also to convert mode 2 (mode 1) to mode 1 (mode 2) are demonstrated by numerical simulations.

## Acknowledgments

The authors acknowledge supports from National Science Council and National Center for Theoretical Sciences (Taipei Office) of Taiwan as well as Center for Quantum Science and Engineering, National Taiwan University (CQSE-10R1004021).